%% file: main.tex
\documentclass[twocolumn, times]{aastex631}

\usepackage{booktabs}
\usepackage{multirow}
\usepackage{amsmath}
\begin{document}
\hypersetup{colorlinks=true, citecolor=blue, urlcolor=blue, linkcolor=blue, allcolors = blue}

\title{Leveraging Machine Learning for Accurate and Fast Stellar Mass Estimation of Galaxies}

\author{Vahid Asadi~\href{https://orcid.org/0009-0005-8897-2385}{\includegraphics[scale=0.04]{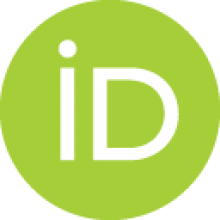}}}
\affiliation{Department of Physics, Institute for Advanced Studies in Basic Sciences (IASBS), PO Box 11365-9161, Zanjan, Iran; \url{vahidasadi@iasbs.ac.ir}}

\author{Akram Hasani Zonoozi~\href{https://orcid.org/0000-0002-0322-9957}{\includegraphics[scale=0.04]{orcid-ID.png}}}
\affiliation{Department of Physics, Institute for Advanced Studies in Basic Sciences (IASBS), PO Box 11365-9161, Zanjan, Iran; \url{vahidasadi@iasbs.ac.ir}}
\affiliation{Helmholtz-Institut f\"ur Strahlen-und Kernphysik (HISKP), Universit\"at Bonn, Nussallee 14-16, D-53115 Bonn, Germany; \url{haghi@iasbs.ac.ir}}

\author{Hosein Haghi~\href{https://orcid.org/0000-0002-9058-9677}{\includegraphics[scale=0.04]{orcid-ID.png}}}
\affiliation{Department of Physics, Institute for Advanced Studies in Basic Sciences (IASBS), PO Box 11365-9161, Zanjan, Iran; \url{vahidasadi@iasbs.ac.ir}}
\affiliation{Helmholtz-Institut f\"ur Strahlen-und Kernphysik (HISKP), Universit\"at Bonn, Nussallee 14-16, D-53115 Bonn, Germany; \url{haghi@iasbs.ac.ir}}
\affiliation{School of Astronomy, Institute for Research in Fundamental Sciences (IPM), PO Box 19395 - 5531, Tehran, Iran}

\author{Fatemeh Abedini~\href{https://orcid.org/0009-0000-5827-5435}{\includegraphics[scale=0.04]{orcid-ID.png}}}
\affiliation{Department of Physics, Institute for Advanced Studies in Basic Sciences (IASBS), PO Box 11365-9161, Zanjan, Iran; \url{vahidasadi@iasbs.ac.ir}}

\author{Atousa Kalantari~\href{https://orcid.org/0009-0006-2285-6792}{\includegraphics[scale=0.04]{orcid-ID.png}}}
\affiliation{Department of Physics, Institute for Advanced Studies in Basic Sciences (IASBS), PO Box 11365-9161, Zanjan, Iran; \url{vahidasadi@iasbs.ac.ir}}

\author{Marziye Jafariyazani~\href{https://orcid.org/0000-0001-8019-6661}{\includegraphics[scale=0.04]{orcid-ID.png}}}
\affiliation{IPAC/Caltech, 1200 E. California Boulevard, Pasadena, CA 91125, USA}
\affiliation{SETI Institute, Mountain View, CA, 94043, USA
}

\author{Nima Chartab~\href{https://orcid.org/0000-0003-3691-937X}{\includegraphics[scale=0.04]{orcid-ID.png}}}
\affiliation{IPAC/Caltech, 1200 E. California Boulevard, Pasadena, CA 91125, USA}



\begin{abstract}
Unveiling the evolutionary history of galaxies necessitates a precise understanding of their physical properties. Traditionally, astronomers achieve this through spectral energy distribution (SED) fitting. However, this approach can be computationally intensive and time-consuming, particularly for large datasets. This study investigates the viability of machine learning (ML) algorithms as an alternative to traditional SED-fitting for estimating stellar masses in galaxies. We compare a diverse range of unsupervised and supervised learning approaches including prominent algorithms such as K-means, HDBSCAN, Parametric t-Distributed Stochastic Neighbor Embedding (Pt-SNE), Principal Component Analysis (PCA), Random Forest, and Self-Organizing Maps (SOM) against the well-established LePhare code, which performs SED-fitting as a benchmark. We train various ML algorithms using simple model SEDs in photometric space, generated with the BC03 code. These trained algorithms are then employed to estimate the stellar masses of galaxies within a subset of the COSMOS survey dataset. The performance of these ML methods is subsequently evaluated and compared with the results obtained from LePhare, focusing on both accuracy and execution time. Our evaluation reveals that ML algorithms can achieve comparable accuracy to LePhare while offering significant speed advantages (1,000 to 100,000 times faster). K-means and HDBSCAN emerge as top performers among our selected ML algorithms. Supervised learning algorithms like Random Forest and manifold learning techniques such as Pt-SNE and SOM also show promising results. These findings suggest that ML algorithms hold significant promise as a viable alternative to traditional SED-fitting methods for estimating the stellar masses of galaxies.

\end{abstract}


\section{Introduction}

A key objective of observational astronomy is to understand the evolution of galaxies over time. This requires deciphering the information encoded in their spectral energy distributions (SEDs) which provide insights into their key physical properties such as stellar mass, star formation rate (SFR), redshifts, age, and dust content.

Assuming the galaxy's star formation history (SFH), the initial mass function (IMF), and how dust affects light as the initial input, one can generate a theoretical SED and compare it to the observed photometric data of the galaxy. The model parameters are then adjusted until the theoretical SED closely matches the observations. Finally, the best-fitting model is used to infer the galaxy's physical properties \citep [e.g.,] [] {Madi2011, Walch2011, Conroy2013, CCH2015}.

Despite their widespread use, traditional galaxy analysis methods based on fitting SEDs are known to introduce uncertainties. These uncertainties arise from the inherent assumptions required about the galaxy's composition, including its SFH, IMF, and dust content. Each of these choices can dramatically impact the derived physical properties. For instance, \cite{Acquaviva2015} evaluated the impact of different modeling assumptions on the recovered SED parameters and found that incorrect parameterizations of the SFH can lead to substantial changes in the derived physical properties. This finding has been corroborated by the works of \cite{Wuy2009}, \cite{Mich2014}, \cite{Sob2014}, and \cite{Simha2014}, all of whom reported significant effects of SFH on the derived galaxy stellar mass across various observational datasets. Additionally, \cite{Pacifici2015} discovered that the simplistic assumption of an exponentially declining SFH, along with a basic dust law and no contribution from emission lines, fails to accurately recover the true star formation rate–mass relationship for star-forming galaxies. Similarly, \cite{Iyer2017} and \cite{Lower2020} found that fitting the SFH using simple functional forms can introduce biases of up to 70\% in the recovered total stellar mass.

\cite{Hemmati2019} demonstrated that Self-Organizing Maps \citep[SOM,][]{Kohonen1981}, trained on a set of SEDs generated using the stellar population synthesis models of \cite{Bruzual2003} (hereafter BC03), can achieve accuracy comparable to traditional SED-fitting for estimating galaxy masses at $z \sim 1$. By learning from extensive and diverse synthetic SED samples, machine learning (ML) models like SOM can efficiently map observed photometry to underlying physical properties of galaxies across a broad and continuous parameter space. In contrast, traditional fitting methods rely on predefined parameter grids, which limit flexibility and may inadequately sample the full parameter space.

The key advantage of ML approaches lies in their ability to capture complex, nonlinear relationships between SEDs and physical properties—relationships that are often difficult to model explicitly using template-based methods. For instance, ML methods can help break classical degeneracies (e.g., age–dust–metallicity) that plague $\chi^2$-based SED fitting. Additionally, ML algorithms exhibit greater robustness to noisy or incomplete data, as they dynamically weight photometric bands based on predictive relevance rather than relying on rigid template matching.

While ML methods inherit some limitations from the synthetic training data-such as simplified SFHs, fixed IMFs, and idealized dust attenuation laws-it mitigates certain biases inherent in traditional approaches and offers significant improvements in computational efficiency.

Moreover, traditional SED fitting can be computationally intensive, which may hinder the analysis of large datasets. This computational challenge is particularly relevant for upcoming large-scale surveys, where efficient analysis methods are critical for timely scientific insights. \cite{Hemmati2019} showed that SOM can achieve speed improvements of up to $10^4$ on a CPU and $10^6$ on a GPU compared to conventional SED-fitting methods. For example, this enables Rubin-LSST's expected 30 billion galaxies to be processed in hours—rather than the several years required by traditional SED-fitting making large-scale survey analysis feasible. Furthermore, advancements in the emulation of expensive SED models, such as neural network-based approaches, offer another promising avenue for mitigating computational costs while retaining accuracy \citep[e.g.,][]{alsing2020speculator,kwon2023neural,mathews2023simple}.

This study expands upon prior work by conducting a comprehensive comparison of multiple ML algorithms—both supervised and unsupervised—for galaxy mass estimation. Unlike previous studies focused on SOM \citep[e.g.,][]{Hemmati2019, Davidzon2022}, we evaluate underexplored techniques, including Parametric t-SNE (Pt-SNE; \citealt{Policar2021}), PCA \citep{mackiewicz1993principal}, K-means \citep{macqueen1967some}, and Random Forest \citep{Breiman2001}. We benchmark their accuracy and computational efficiency against traditional SED-fitting using \textsc{LePhare} \citep{Arnouts1999, Ilbert2006}. Our goals are twofold: (1) to demonstrate the broad applicability of ML algorithms for galaxy mass estimation, and (2) to identify the optimal balance of accuracy and speed relative to SED-fitting.

This paper is organized as follows: Sect.~\ref{sec:2} describes the data used in this study. Sect.~\ref{sec:3} provides a concise overview of the algorithms employed. Sect.~\ref{sec:4} explains the methods used for stellar mass estimation. Sect.~\ref{sec:5} delves into a comparison of the performance of different trained models. Finally, Sect.~\ref{sec:6} discusses our most successful models and potential future directions.

We adopt a flat $\Lambda$CDM cosmology with parameters $H_{0}=70kms^{-1}Mpc^{-1}$, $\Omega_{m}=0.3$ and $\Omega_{\Lambda}=0.7$. All magnitudes are reported in the AB system \citep{Oke1974}.

\section{Data}\label{sec:2}

\subsection{COSMOS galaxies}
In our comparative analysis of ML algorithms, we utilize broadband photometry data from the COSMOS2015 survey \citep{Sovi2007, Laigle2016}. This catalog includes apparent magnitudes across 30 bands, spanning from ultraviolet (UV) to infrared (IR) wavelengths (0.25–8 µm). In the optical range, it features six broad bands: B, V, r, i, $z^{++}$, along with 12 medium bands and 2 narrow bands, all captured using the Subaru Suprime-Cam \citep{taniguchi2007cosmic}. The $u^{*}$-band data are obtained from the Canada–Hawaii–France Telescope (CFHT/MegaCam). In the near-infrared (NIR), the catalog relies on Y, J, H, $K_{s}$ images from the UltraVISTA survey \citep{mccracken2012ultravista}. Additionally, mid-infrared data from the four IRAC channels (covering a wavelength range of approximately 3 to 8 µm) are obtained from the SPLASH program (PI: Capak). Photometric redshifts, stellar masses, absolute magnitudes, and star formation rates (SFRs) are derived using the LePhare code.

\begin{figure}
        \centering
        \includegraphics[width=0.84\linewidth]{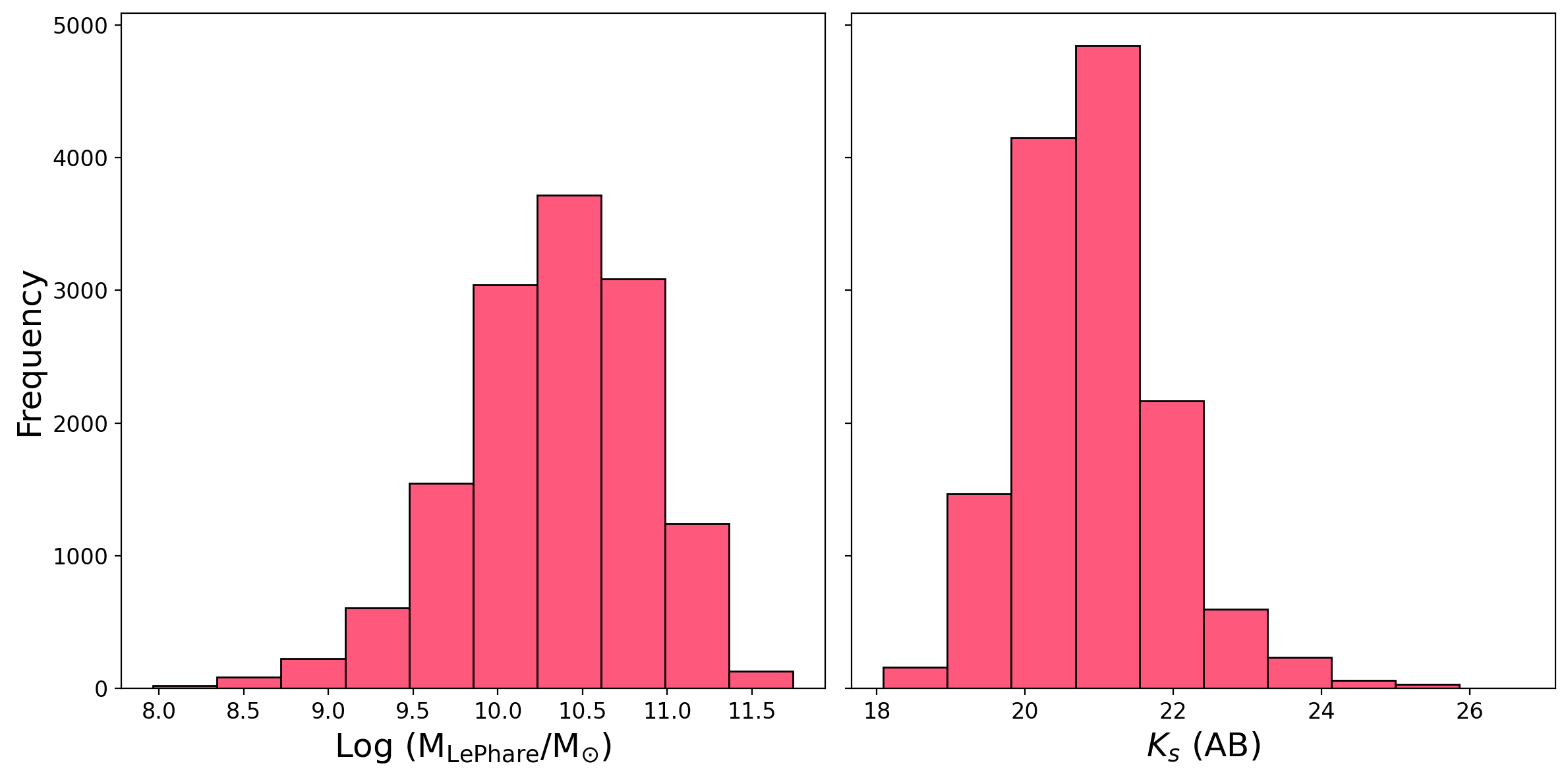}
        \caption{Histograms showing the distributions of stellar mass and $K_{s}$-band magnitude for a sample of approximately 14,000 galaxies selected from the COSMOS2015 survey. The sample is limited to galaxies with LePhare redshifts between 0.8 and 1.2.}
        \label{fig:fig0}
\end{figure}

For this study, a well-defined subsample of $\sim$14,000 galaxies from COSMOS2015 is analyzed, using data from 12 photometric bands (see Table~\ref{tab:tab1}). To ensure robust mid-infrared detection, a magnitude cut of $m_{\text{ch1}} < 26$ is applied to exclude faint sources. Stellar sources and X-ray-detected objects are removed to obtain a pure galaxy sample. The analysis is further restricted to galaxies within the redshift range $0.8 < z < 1.2$ (as derived by LePhare), and objects with missing values in any of the selected bands are excluded. Figure~\ref{fig:fig0} shows the LePhare-derived stellar mass and $K_s$-band magnitude distributions for the sample.

\subsection{BC03 synthetic SEDs}
To train the ML algorithms, we utilized the widely used stellar population synthesis code developed by \citet{Bruzual2003} (hereafter BC03) to generate galaxy spectra with different physical properties. The SED of modeled galaxies is constructed by combining simple stellar populations with a range of stellar ages ($7.7<\log_{10}(\rm age/yr)<10.0$), according to an exponentially declining SFH. The e-folding timescale, $\tau$, is adopted to vary in the range of 0.1-10 Gyr. For each simple stellar population, a sub-solar metallicity ($0.4Z_{\odot}$) is adopted and stars are assumed to form following the Chabrier IMF \citep{Chap2003}. Furthermore, we added Calzetti extinction \citep{Calzetti2000} with varying degrees of reddening ($0<E(B-V)<1$) to the synthetic SEDs.

To generate synthetic photometry that can be compared to our observed data, we adjusted the models to account for a redshift of approximately $z \approx 1$ and integrated them using the same filter transmissions as in the COSMOS observations (see Table~\ref{tab:tab1}). As a result, we produced around 14,000 models.

\input{band_table}

\section{Algorithms}\label{sec:3}
In this section, we describe the ML algorithms used in our study to predict stellar mass from synthetic galaxy colors. Our primary objective is to evaluate how effectively different classes of ML algorithms can recover stellar mass based on the shape of the SED, encoded as broad-band photometric colors. We assess their performance in terms of accuracy (relative to SED-fitting) and computational efficiency. We compare four distinct ML approaches for this task: (1) manifold learning algorithms, (2) projection algorithms, (3) clustering algorithms, and (4) regression algorithms (summarized in Table~\ref{tab:tab2}). By training these algorithms on synthetic data generated from stellar population models, we evaluate their ability to capture the underlying physics encoded in the color--mass relation. This section provides a concise overview of these methods, while Section~\ref{sec:4} details their application to stellar mass estimation.

\subsection{Manifold learning}
Manifold learning \citep[e.g.,] []{izenman2012,meilua2024manifold} is a dimensionality reduction technique that operates under the assumption that high-dimensional data often lies on a lower-dimensional, hidden structure known as a manifold. Manifold learning algorithms seek to uncover this lower-dimensional representation using various approaches. In this section, we will explain the manifold learning algorithms considered in this study: SOM, Pt-SNE, UMAP\footnote{Uniform Manifold Approximation and Projection} and ISOMAP\footnote{Isometric Mapping}.

\subsubsection{SOM}
SOM algorithm \citep{Kohonen1981} reveals the hidden patterns in high-dimensional data by utilizing a grid of interconnected nodes. Each node possesses a weight vector, serving as an anchor point on the data's high-dimensional manifold. Through competitive learning, data points compete to activate the closest node (winner), which, along with its neighbors, adjusts its weights to resemble the data point. This process enables the SOM to learn the local and global structure of the manifold. Once trained, the SOM can project new, unseen data points onto its compressed manifold, estimating their location within the manifold and providing valuable insights into their relationships with previously seen data.

\subsubsection{Pt-SNE} 
Pt-SNE \citep{Policar2021} addresses the limitation of traditional t-SNE \citep{van2008visualizing} by learning a parametric mapping function, typically a neural network $f(x; \theta)$, that transforms high-dimensional data points x to low-dimensional embeddings y. Like t-SNE, Pt-SNE models pairwise similarities using probability distributions: Gaussian distributions in the high-dimensional space and Student's t-distributions in the low-dimensional space. The network parameters $\theta$ are optimized by minimizing the Kullback-Leibler (KL) divergence between these distributions. This learned function enables the projection of new, unseen data points onto the low-dimensional embedding, a capability absent in standard t-SNE, making Pt-SNE more practical for visualizing and analyzing evolving datasets.

\input{uni_table0}

\subsubsection{UMAP} 
UMAP \citep{mcinnes2018umap} construct fuzzy "neighborhood graphs" based on the proximity of data points, akin to interconnected webs where connections signify the closeness of data points in the high-dimensional space. UMAP's strength lies in optimizing a similar graph in the lower-dimensional space, striving to perfectly replicate the connections of the original. By meticulously preserving these connections, UMAP captures both the local nuances and the broader structure of the data, allowing unseen data points to be seamlessly projected onto this new map. This reveals intricate relationships and provides a clearer view of the complex data's hidden patterns and structures.

\subsubsection{ISOMAP} 
ISOMAP \citep{Tenenbaum2000AGG} addresses the challenge of visualizing high-dimensional data by uncovering its intrinsic geometric structure. Unlike Euclidean distance, which measures straight-line separation, ISOMAP focuses on geodesic distances, the true distances along a potentially warped, lower-dimensional manifold where the data resides. By constructing a neighborhood graph connecting nearby data points, ISOMAP effectively mimics the underlying manifold, capturing its intricate geometry. It then calculates the shortest paths (geodesics) between all points on this graph, essentially tracing the manifold's curvature. Finally, multidimensional scaling is employed to find a lower-dimensional representation that best reflects these geodesic distances, unfolding the complex manifold and revealing the true relationships between data points. This process enables ISOMAP to excel at capturing the global structure of the data, offering valuable insights into the organization of complex datasets and allowing even unseen data points to be seamlessly projected onto the map.

\subsection{Projection}
Unlike manifold learning, we don't necessarily assume the data lies on a specific low-dimensional structure in projection-based dimensionality reduction. Instead, we use linear or non-linear transformations to project the data onto a lower-dimensional space while attempting to preserve important information from the original data. PCA \citep[e.g.,] []{mackiewicz1993principal,abdi2010principal,greenacre2022principal} is the most prominent. Unlike distance-preserving techniques, PCA focuses on intrinsic data variation by identifying the directions of maximum variance—the principal components—which often reflect the underlying manifold structure of the data. By projecting the data onto these components, PCA constructs a lower-dimensional representation that retains the most informative features. This approach not only captures essential patterns but also generalizes to unseen data: new points can be consistently mapped into the same reduced space, maintaining their relationships with the original dataset.

\subsection{Clustering}
Clustering in unsupervised learning \citep[e.g., see ] [] {saxena2017review} is the process of grouping unlabeled data points based on their similarities or differences. Clustering algorithms discover intrinsic patterns and structures within the data, revealing clusters of related data points without any pre-defined labels or guidance. In the following, we discuss two prominent clustering algorithms: K-means and HDBSCAN\footnote{Hierarchical Density-Based Spatial Clustering of Applications with Noise}.

\subsubsection{K-means} 
K-means \citep[e.g.,] []{macqueen1967some, jain2010data} operates iteratively, employing a cyclical process to cluster data. Initially, k centroids (cluster centers) are randomly selected from the data. Then, each data point is assigned to the nearest centroid based on a distance metric, such as Euclidean distance. Next, the centroids are recalculated as the mean of their assigned points. These steps - assignment and re-centroiding - are repeated until the centroids converge, indicating stability. While K-means is effective, it is sensitive to the initial placement of centroids and requires prior specification of the number of clusters (k).

\subsubsection{HDBSCAN} 
HDBSCAN \citep{mcinnes2017hdbscan} tackles unlabeled data, uncovering clusters of various shapes and densities without needing pre-defined cluster counts. It extends the traditional DBSCAN \citep{ester1996density} algorithm, which identifies clusters based on density but requires a fixed distance parameter (epsilon) and can struggle with varying cluster densities. Unlike DBSCAN, HDBSCAN leverages the data's inherent density to identify meaningful clusters by finding high-density areas (core samples) and analyzing how they connect through border points.

HDBSCAN meticulously examines both core density and reachability distances, building a hierarchical structure that allows it to extract clusters at varying densities. This capability enables it to effectively handle even low-density or noise-obscured clusters, making it more robust to parameter selection compared to DBSCAN, which can be sensitive to the choice of epsilon.

\subsection{Regression techniques}
Regression tasks in machine learning aim to uncover the functional relationship between independent variables (features) and a continuous dependent variable (target). The model learns this association by fitting a function to the training data. This function can then be used to predict target values for new, unseen data points based solely on their features. In this section, we briefly discuss Random Forest and K-Nearest Neighbors (KNN) algorithms considered in this study.

\subsubsection{Random Forest regression} 
Random forest regression \citep[e.g.,] []{Breiman2001, Geurts2006} is a robust predictive modeling technique that leverages the strength of ensemble methods to tackle complex prediction tasks in high-dimensional data. This approach operates by constructing a multitude of decision trees, each trained on a random subset of features and data points (bootstrapping). When making predictions, the forest aggregates the predictions from each tree, effectively creating a democratic voting system. This ensemble approach significantly reduces variance and improves prediction accuracy compared to a single decision tree, making it a powerful tool for predictive modeling.

\subsubsection{KNN} 
KNN \citep{peterson2009k} regression is a simple yet effective predictive modeling technique that harnesses the similarity between data points to tackle prediction tasks. When faced with a new, unseen data point, KNN regression identifies its k nearest neighbors within the training data based on a distance metric, such as Euclidean distance. The predicted value for the unseen point is then estimated by averaging the corresponding target values of its k neighbors. 

\begin{figure*}
        \centering
        \includegraphics[width=\linewidth]{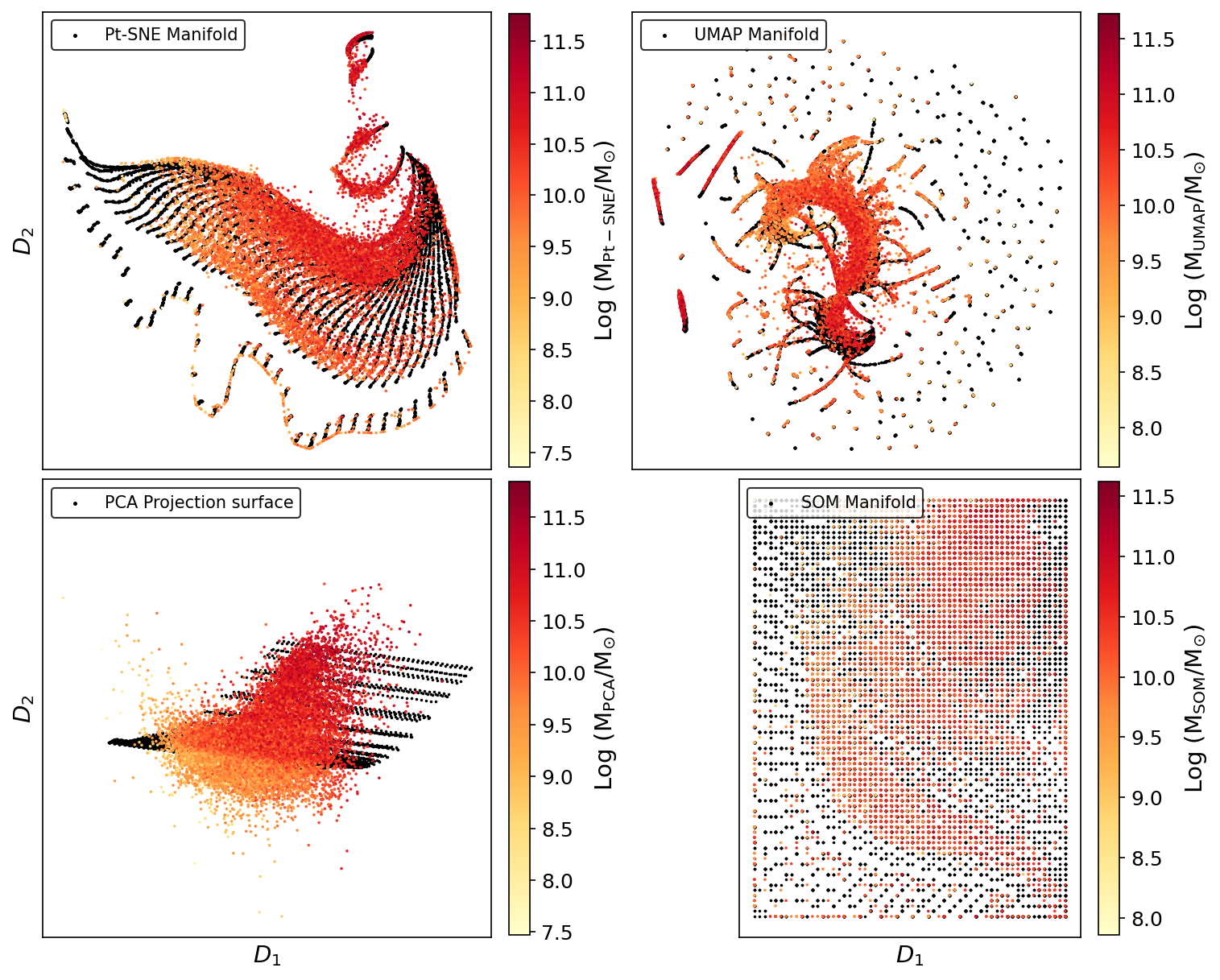}
        \caption{Manifolds of Pt-SNE, UMAP and SOM, along with the projection surface of PCA (black points), with the mapped COSMOS galaxies (colored points). Galaxy colors correspond to their estimated stellar masses obtained with the mentioned algorithms. The dimensions of the 2D plots (D1 and D2) are arbitrary labels and do not carry any physical significance.}
        \label{fig:fig1}
\end{figure*}

\section{Methods}\label{sec:4}
To estimate the stellar mass of galaxies, we trained all algorithms listed in Table~\ref{tab:tab2} using a sample of synthetic galaxies generated by BC03. In this study, the training parameter space has 11 dimensions representing the colors of galaxies. These colors are constructed based on the broadband photometry listed in Table~\ref{tab:tab1}, with the following pairings: $u^* - B, B - V, V - r, r - i^{+}, i^{+} - z^{++}, z^{++} - Y, Y - J, J - H, H - K_{s}, K_{s} - ch1, ch1 - ch2$.

We opted to use color indices rather than apparent magnitudes because they provide a more direct and robust probe of the physical properties of galaxies. Unlike apparent magnitudes, which are sensitive to distance-dependent effects and flux calibration uncertainties, color indices are distance-independent and more resilient to systematic biases. Additionally, because colors measure the relative flux across different wavelengths, they better constrain the shape of the SED, which is crucial for distinguishing between different stellar populations and dust attenuation effects.

For each synthetic galaxy, we adopted the median mass-to-light ratio in the $K_{s}$ band computed by BC03 code. ML algorithms were trained to predict this ratio from galaxy colors. For an observed galaxy, the model outputs an estimated M/$L_{K_{s}}$, which is then multiplied by the observed $K_{s}$-band luminosity to derive the stellar mass:

\begin{equation}
M_{*} = \left( \frac{M}{L_{K_{s}}} \right)_{ML-preadicted} \times L_{K,\text{observed}}
\end{equation}

Prior to training, we scaled both the synthetic galaxies and COSMOS galaxies using the \texttt{StandardScaler} function\footnote{\url{https://scikit-learn.org}} which performs standardization by transforming each color feature to have zero mean and unit variance:

\begin{equation}
X_{\text{scaled}} = \frac{X - \mu}{\sigma},
\end{equation}
where $\mu$ is the mean and $\sigma$ is the standard deviation of each color feature calculated from galaxy data. This preprocessing step ensures all color dimensions contribute equally during machine learning by removing arbitrary differences in scale between colors.

With the exception of K-means, we used default settings for all algorithms (see source links in Table~\ref{tab:tab2}). This approach was motivated by preliminary tests showing that optimized hyperparameters did not significantly improve performance over default configurations, suggesting the defaults were already well-suited to our dataset. Our empirical validation revealed that parameter tuning (e.g., number of estimators in Random Forest, number of neighbors in KNN, or perplexity in Pt-SNE) typically yielded improvements smaller than 0.01\,dex in $\sigma_{F}$ (Equation~\ref{eq:rms}), confirming the adequacy of default configurations. Furthermore, modern ML libraries implement carefully chosen defaults that perform robustly across diverse datasets.

In the following sections, we will explain in detail how we estimated the stellar mass of galaxies using different categories of ML algorithms. 

\subsection{Manifold and Projection algorithms}
With the exception of SOM, the training process algorithms for these two categories are similar. All algorithms were trained on the color data of synthetic galaxies and then reduced to two dimensions. For SOM, a $60 \times 80$ grid was used during training. Subsequently, COSMOS galaxies were mapped onto the manifold or projection surfaces learned by the algorithms.

To estimate galaxy masses, the mass-to-light ratio label associated with each cell was used for the SOM. For the other algorithms, we employed a distance-weighted mean of the mass-to-light ratio labels for the five nearest training data points. After assigning a mass-to-light ratio to each mapped galaxy, its luminosity was used for mass estimation. The manifolds of Pt-SNE, UMAP, and SOM, and the projection surface of PCA, with mapped galaxies, are illustrated in Figure~\ref{fig:fig1}.

\subsection{Clustering algorithms}
We employed K-means clustering to identify potential galaxy clusters within the BC03 synthetic galaxies. By setting the number of clusters (K) to 20, we aimed to find the most representative synthetic galaxy for each cluster. These representative galaxies are the ones closest to their respective cluster centroids as measured by Euclidean distance in the 11-dimensional color space.

To estimate the mass of a typical galaxy, we compared its 11-dimensional color data to that of the 20 representative galaxies. We selected the representative galaxy with the most similar color data and used its corresponding mass-to-light ratio. By multiplying this ratio by the galaxy's luminosity, we calculated an estimate of its mass.

\begin{figure}
    \centering
    \includegraphics[width=\linewidth]{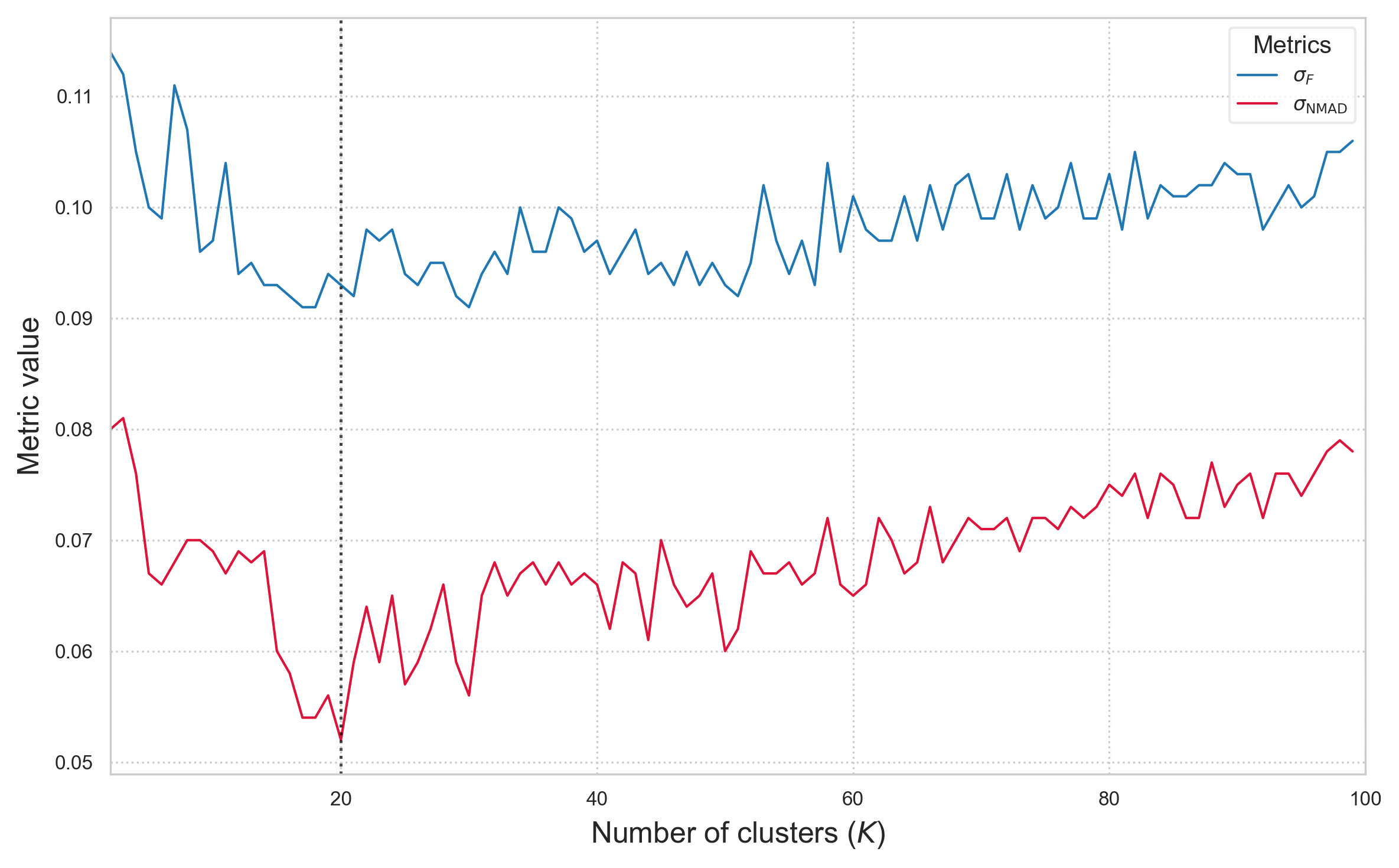}
    \caption{Optimization of the number of K-means clusters (K). The root-mean-square error ($\sigma_{F}$) and normalized median absolute deviation ($\sigma_{NMAD}$) between the estimated and true mass-to-light ratios of the test set galaxies are plotted as a function of K. The minimum values observed around K=20 suggest this as the optimal number of clusters for our analysis.}
    \label{fig:figkmeans}
\end{figure}

To determine the optimal number of clusters (K), we randomly partitioned the model galaxies into a training set (70\%) and a test set (30\%). We varied K from 2 to 100, identifying representative galaxies for each value. The test set galaxies were then matched to these representatives, and their mass-to-light ratios were estimated. By comparing these estimated mass-to-light ratios to the true test set values using the root-mean-square error (Equation~\ref{eq:rms}) and normalized median absolute deviation (Equation~\ref{eq:nmad}), we found that K = 20 provided the optimal clustering solution, as both metrics reached approximate minima at this value (see Figure~\ref{fig:figkmeans}).

Since K-means is sensitive to initial centroid placement \citep[e.g.,][]{celebi2013comparative}, we mitigated initialization instability by using the final centroids of the 20 synthetic clusters as the initial centroids when applying the algorithm to real galaxies.

Similarly, we applied HDBSCAN clustering, which yielded 664 clusters. HDBSCAN does not require pre-specifying the number of clusters and can identify clusters of varying densities. To estimate stellar masses, we identified the medoids (the data points closest to the center of each cluster). For each COSMOS galaxy in our sample, we then found the most similar medoid based on color data. The mass-to-light ratio associated with the medoid was then used, along with the luminosity of the COSMOS galaxy, to estimate its stellar mass through multiplication, similar to the K-means approach.

\subsection{Regression algorithms}
We trained all regression algorithms in Table~\ref{tab:tab2} on the color data of BC03 synthetic galaxies as input features, using their corresponding mass-to-light ratios as target variables. This allows us to estimate the mass of a COSMOS galaxy by multiplying its predicted mass-to-light ratio by its known luminosity.
\input{uni_table}

\section{Results}\label{sec:5}
This section evaluates the accuracy and efficiency of the ML method in comparison to the LePhare method for estimating the stellar masses of COSMOS galaxies. The ML approach employs algorithms trained on the color data of BC03 synthetic galaxies, as described in Sect.~\ref{sec:4}. In contrast, LePhare fits BC03 synthetic magnitudes to the observed magnitudes of COSMOS galaxies across the selected bands. We compare the performance of the ML models to LePhare in terms of both accuracy and computational speed, using the latter as a benchmark. Additionally, we use these models to estimate the error bars associated with the galaxy mass estimates and perform a subsequent comparison.

\begin{figure*}
        \centering
        \includegraphics[width=\linewidth]{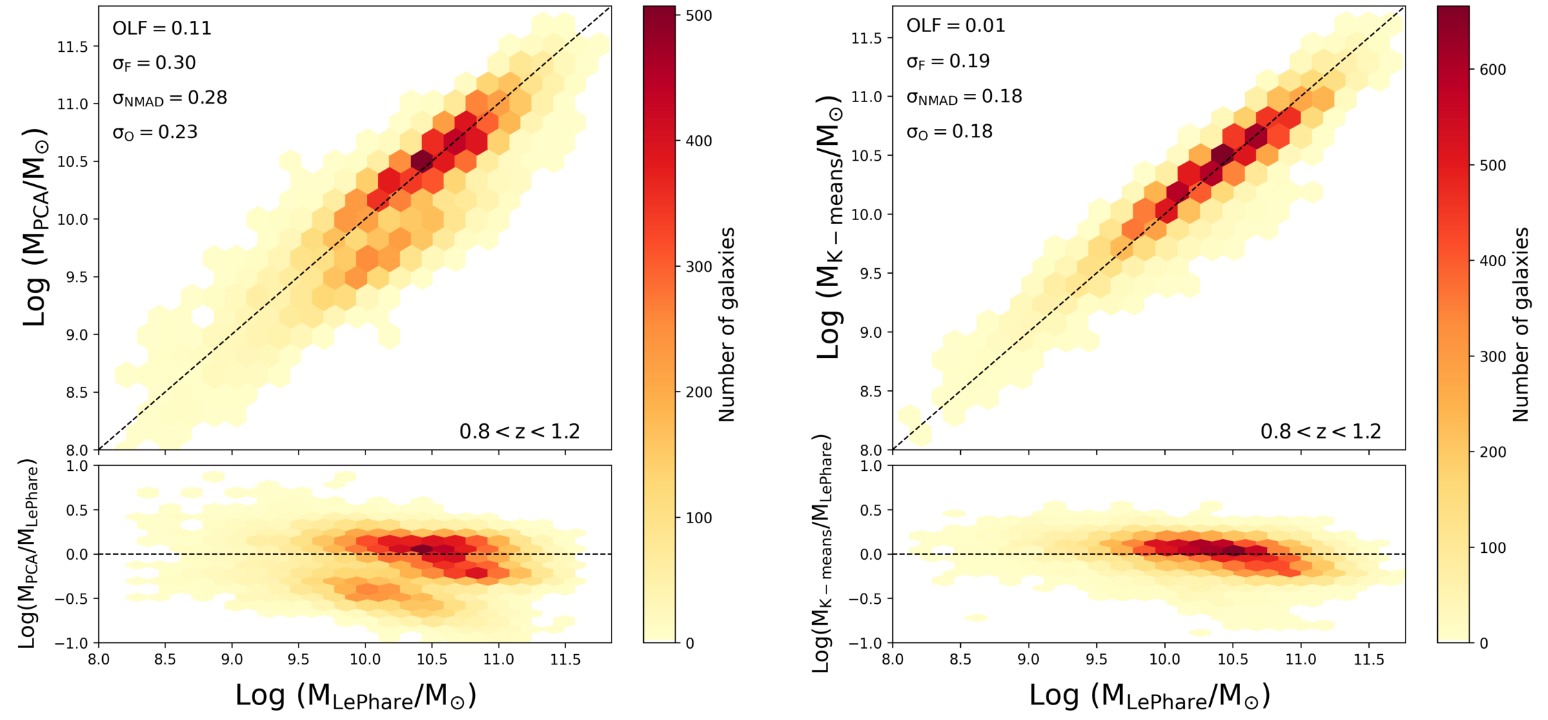}
        \caption{Stellar masses of COSMOS galaxies measured by PCA and K-means compared to those derived from LePhare.}
        \label{fig:fig3}
\end{figure*}

\subsection{Stellar mass estimates}
\subsubsection{Accuracy}
We evaluated the accuracy of the ML algorithms using two metrics \citep{Nayerii2017}: the root-mean-square (RMS) of the logarithmic difference between the algorithm's mass predictions ($M_{Algorithm}$) and the LePhare predictions ($M_{LePhare}$) denoted by $\sigma_{F}$, and the normalized median absolute deviation (NMAD) of the logarithmic difference, denoted by $\sigma_{NMAD}$.

\begin{equation}
\sigma_{F}=rms(log(M_{Algorithm})-log(M_{LePhare}))
\label{eq:rms}
\end{equation}

\begin{equation}
\sigma_{NMAD} = 1.48 \times median(|log(M_{Algorithm})-log(M_{LePhare})|)
\label{eq:nmad}
\end{equation}

We then identified and removed outliers based on the outlier fraction (OLF). The OLF is defined as the proportion of objects where the absolute value of the difference in log-transformed mass between the algorithm's prediction and the LePhare prediction is greater than 0.5 dex, corresponding to a 3-$\sigma$ deviation \citep{mobasher2015critical}. After outlier removal, we recalculated the RMS of the log-transformed mass difference, denoted by $\sigma_{0}$.

Table~\ref{tab:tab3} provides a summary of the comparison between stellar masses estimated by unsupervised and supervised algorithms and those derived from LePhare. Overall, the majority of the algorithms demonstrate good performance with respect to LePhare. 

Diving deeper into the manifold learning algorithms, a clear distinction emerges. ISOMAP exhibits a significantly higher OLF and $\sigma_{F}$, suggesting it might capture a data structure fundamentally different from the other algorithms. Focusing on the remaining algorithms (Pt-SNE, SOM and UMAP), we see a more balanced picture. They all demonstrate comparable performance across the metrics, indicating a similar ability to capture the underlying manifold structure while remaining consistent with the LePhare method.

\begin{figure*}
        \centering
        \includegraphics[width=\linewidth]{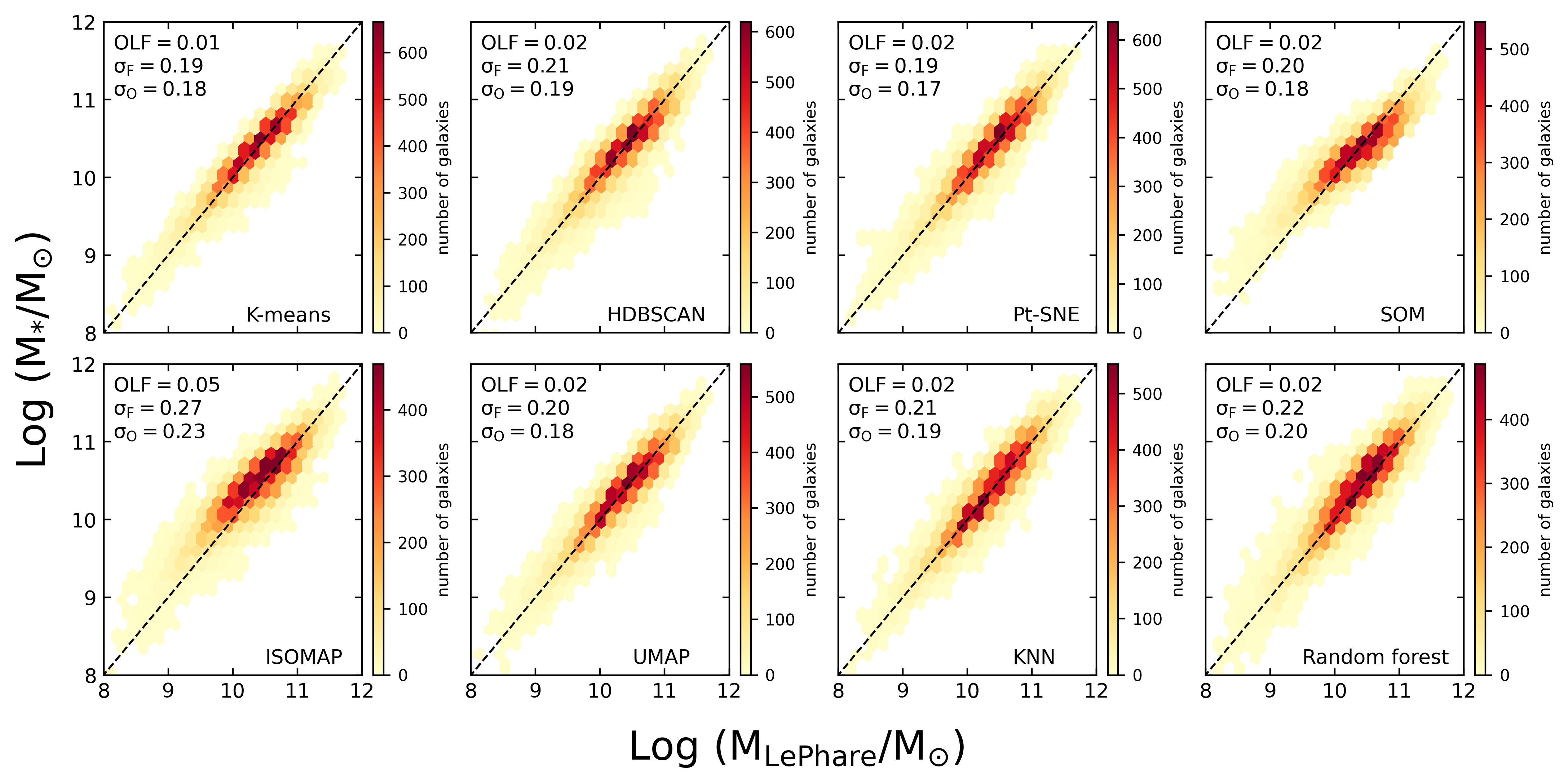}
        \caption{Stellar masses of COSMOS galaxies estimated using eight ML algorithms (listed in Table~\ref{tab:tab3}) compared to those derived from LePhare.}
        \label{fig:fig4}
\end{figure*}

Analysis of PCA's performance in Table~\ref{tab:tab3} reveals significant differences compared to other algorithms, in OLF and variation metrics ($\sigma_{F}$ and $\sigma_{NMAD}$). PCA demonstrates notably higher values for these metrics, indicating poorer performance. The left panel of Figure~\ref{fig:fig3} further supports this finding, showing a distinct bimodal distribution when comparing PCA-predicted masses with LePhare estimates. This bimodality likely stems from nonlinear structure in the underlying data - an inherent limitation for PCA, which relies on linear transformations and consequently may fail to capture important nonlinear relationships between data features.

Compared to PCA, manifold learning algorithms often outperform them. One possible reason for this is their focus on uncovering the data's inherent, lower-dimensional structure, even when it resides on a nonlinear manifold within a higher-dimensional space. This emphasis on intrinsic structure, rather than simply maximizing variance, leads to more faithful representations of complex, nonlinear data.

For clustering algorithms, both K-means and HDBSCAN exhibit low OLF with minimal variations. However, K-means demonstrates a slight edge in terms of OLF. It achieved the lowest OLF among all algorithms compared in this study. The right panel of Figure~\ref{fig:fig3} compares the stellar masses estimated by K-means for the COSMOS galaxies with those derived from LePhare.

Our analysis of Table~\ref{tab:tab3} shows that both KNN and Random Forest perform comparably to manifold learning and clustering algorithms. Their success likely stems from their non-parametric nature, as neither method makes assumptions about the underlying data distribution. This flexibility enables them to effectively capture complex, non-linear patterns in the data. KNN operates by predicting values based on feature-space similarity, effectively learning decision boundaries directly from the data. Random Forest, meanwhile, aggregates predictions from multiple decision trees, each capable of modeling non-linear relationships through hierarchical branching conditions.

Figure~\ref{fig:fig4} compares stellar mass estimates obtained using eight ML algorithms from Table~\ref{tab:tab3} for the COSMOS galaxies with those derived from LePhare.

To evaluate the resilience of our ML algorithms to photometric uncertainties, we introduced Gaussian noise ($\sigma = 0.05$\,mag) to the synthetic data. The Random Forest algorithm demonstrated exceptional noise resistance, showing nearly identical performance ($\sigma_{F} = 0.223$ vs $0.220$ noiseless)-the smallest variation among all tested methods. Other algorithms also showed minimal changes: Pt-SNE ($0.214$ vs $0.193$), UMAP ($0.21$ vs $0.201$), ISOMAP ($0.238$ vs $0.268$), SOM ($0.226$ vs $0.202$), K-means ($0.21$ vs $0.190$), HDBSCAN ($0.204$ vs $0.213$), PCA ($0.24$ vs $0.303$), and KNN ($0.192$ vs $0.206$). All variations remained below $0.06$\,dex, with PCA showing the largest change while other methods stayed within $\pm0.03$\,dex of their noiseless performance, demonstrating their resilience to photometric uncertainties.

Although this study focuses on comparing ML predictions with LePhare \citep{Hemmati2019, Davidzon2022}, we acknowledge that this approach does not provide validation against an absolute ground truth. Our primary goal here is to assess the relative performance of ML methods within the context of standard analysis pipelines. However, absolute validation remains crucial. To this end, we are conducting a follow-up study (Asadi et al., in preparation) in which both ML and SED-fitting techniques, are benchmarked against synthetic galaxy catalogs derived from cosmological simulations with known stellar masses.

\begin{figure*}
        \centering
        \includegraphics[width=\linewidth]{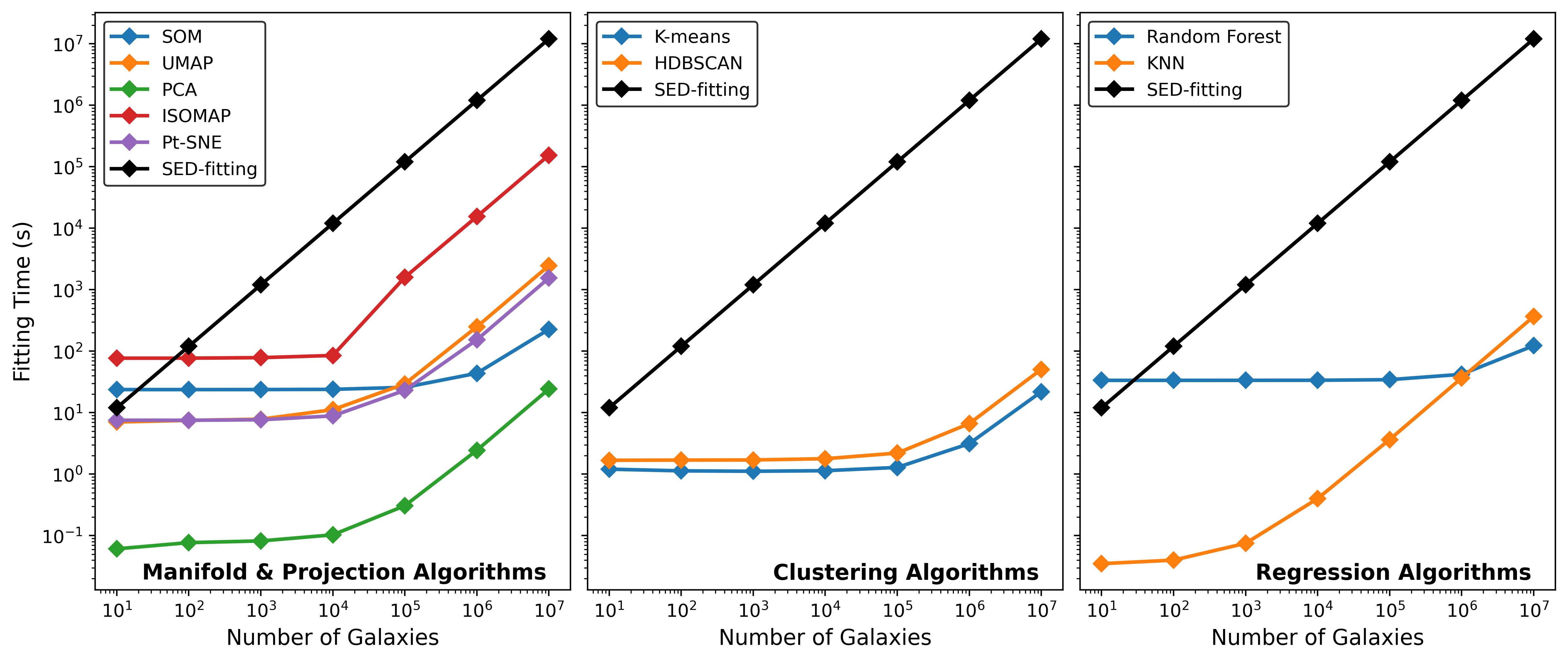}
        \caption{This figure illustrates the comparative runtimes of various ML algorithms across different categories (manifold learning, projection, clustering and regression), alongside SED-fitting with LePhare. The ML algorithms were trained on the color data of synthetic galaxies to estimate the stellar masses of different numbers of real galaxies.}
        \label{fig:fig5}
\end{figure*}

\subsubsection{Computational speed}
Having compared the accuracy of various ML algorithms in this study, we now shift our focus to their computational efficiency. Our goal is to compare the computational performance of the ML algorithms considered in this study both against the LePhare method and against one another. To achieve this, we employed a machine equipped with an 11th Gen Intel® Core™ i7-1165G7 processor to evaluate the computational speed of these algorithms and the LePhare method on a dataset of 10 million galaxies which is constructed synthetically by replicating our original sample of galaxies.

Figure~\ref{fig:fig5} compares the runtimes of all studied algorithms, including SED-fitting with LePhare, across different samples of galaxies categorized into three groups: manifold learning and projection algorithms, clustering algorithms, and regression algorithms. Additionally, Table~\ref{tab:tab3} summarizes the average speedup of these algorithms relative to SED-fitting.

The results indicate that SED-fitting is significantly slower than ML approaches. This slowdown occurs because SED-fitting analyzes each galaxy independently, leading to linear scaling with the dataset size; in other words, the runtime increases proportionally with the number of galaxies. In contrast, ML algorithms require training only once using a small sample of synthetic galaxies. After this initial training, we can perform mapping, comparison, or prediction processes for estimating the stellar masses of real galaxies, as discussed in Sect.~\ref{sec:4}. These efficient processes significantly reduce computation time compared to SED-fitting.

While all ML categories outperform SED-fitting, further analysis reveals efficiency variations within them. PCA emerges as the fastest algorithm, followed by clustering algorithms, which demonstrate the next best overall performance. Regression algorithms are subsequently faster than manifold learning algorithms.

The runtime of manifold learning and projection algorithms encompasses several stages: training with synthetic galaxies, mapping real galaxies and identifying the nearest data points (neighbors) in the trained manifold or projection surface for the mapped galaxies (used for estimating galaxy masses). The latter time is consistent across all algorithms except SOM, which is faster. This is because SOM compares mapped galaxies with a set of 4,800 weight points (arranged in a $60 \times 80$ grid) instead of the 14,000 data points used for training in other algorithms. For smaller galaxy samples (up to $10^4$), training time dominates the overall runtime. However, as the number of galaxies increases significantly ($10^5$ to $10^7$), the mapping time becomes the dominant factor.

Within the manifold learning algorithms, Pt-SNE and UMAP generally exhibit faster training times compared to others. However, SOM has significantly faster mapping times. ISOMAP shows the poorest performance in both training and mapping. This is a consequence of its reliance on all-pairwise distances and eigendecomposition. Calculating distances between every trained data point makes ISOMAP significantly slower than methods that focus on local neighborhoods. Additionally, When galaxies are mapped on the ISOMAP manifold, their mapping time significantly increases. This is because both computationally expensive operations require recalculating the shortest paths and updating the distance matrix \citep{balasubramanian2002isomap}.

PCA, a projection algorithm, excels in computational efficiency for both training and mapping from manifold learning algorithms. Its reliance on linear operations allows for the rapid computation of eigenvectors and eigenvalues, resulting in a swift training process. Mapping galaxies also involves a straightforward projection, keeping the computational cost low. However, it is worth noting that replacing PCA's linear kernel with a nonlinear alternative (e.g., polynomial or sigmoid) can improve accuracy by capturing nonlinear relationships—but at the cost of significantly longer training and mapping times. This trade-off arises primarily from the kernel matrix, which encodes relationships between training data points based on the chosen kernel function. While computing the kernel matrix is computationally expensive, the greater challenge occurs when mapping galaxies onto the trained projection. Unlike linear PCA, which projects galaxies directly onto the precomputed linear space, nonlinear PCA requires evaluating the kernel function between each new galaxy and every training data point used to construct the original projection. This operation scales linearly with the number of mapped galaxies \citep{sriperumbudur2022approximate}, making it far more demanding than the linear case.

For clustering algorithms, the runtime involves three steps: training, identifying the best representative data point in each cluster (as detailed in Sect.~\ref{sec:4}), and finding the nearest representative data points for each real galaxy. Both K-means and HDBSCAN are exceptionally fast, even with a large number of galaxies (as shown in Figure~\ref{fig:fig5}). In fact, they are the fastest algorithms among all those considered in this study (see Table~\ref{tab:tab3}). However, K-means outperforms HDBSCAN due to the larger number of representative data points used by HDBSCAN (664 compared to 20 for K-means).

The speed of regression algorithms can be evaluated based on their training and prediction times. As shown in Figure~\ref{fig:fig5} (Regression section), Random Forest and KNN exhibit contrasting behavior. While its training time is longer compared to other algorithms, it scales well, maintaining a relatively constant runtime as the number of galaxies increases. This results in significantly fast prediction times, as Random Forest prediction relies on traversing pre-built decision trees, a process that is independent of the number of galaxies being analyzed \citep{Geurts2006}. In contrast, KNN boasts a fast training time but suffers from the slowest prediction time. This contrasting behavior stems from their underlying mechanisms. KNN classifies galaxies by identifying the nearest neighbors within the training data. During training, KNN simply stores this data. However, at prediction time, KNN must compute the distance between each galaxy and every trained data point in the training set to identify the nearest neighbors \citep{peterson2009k}. This computation significantly increases prediction time as the number of galaxies grows.

It is worth mentioning that batch processing can mitigate memory limitations and computational demands for some algorithms. For example, using a batch size of 100 galaxies increases the average speed of ISOMAP relative to LePhare by approximately 10 times (from $5.5 \times 10^{1}$ to $5.3 \times 10^{2}$). However, while batch processing helps, it cannot fully compensate for fundamental differences in algorithmic efficiency. For instance, using a batch size of 100 for Random Forest with 10 million galaxies results in a runtime of approximately 700 seconds, whereas its efficient prediction process achieves a runtime of just about 120 seconds.

\begin{figure}
    \centering
    \includegraphics[width=\linewidth]{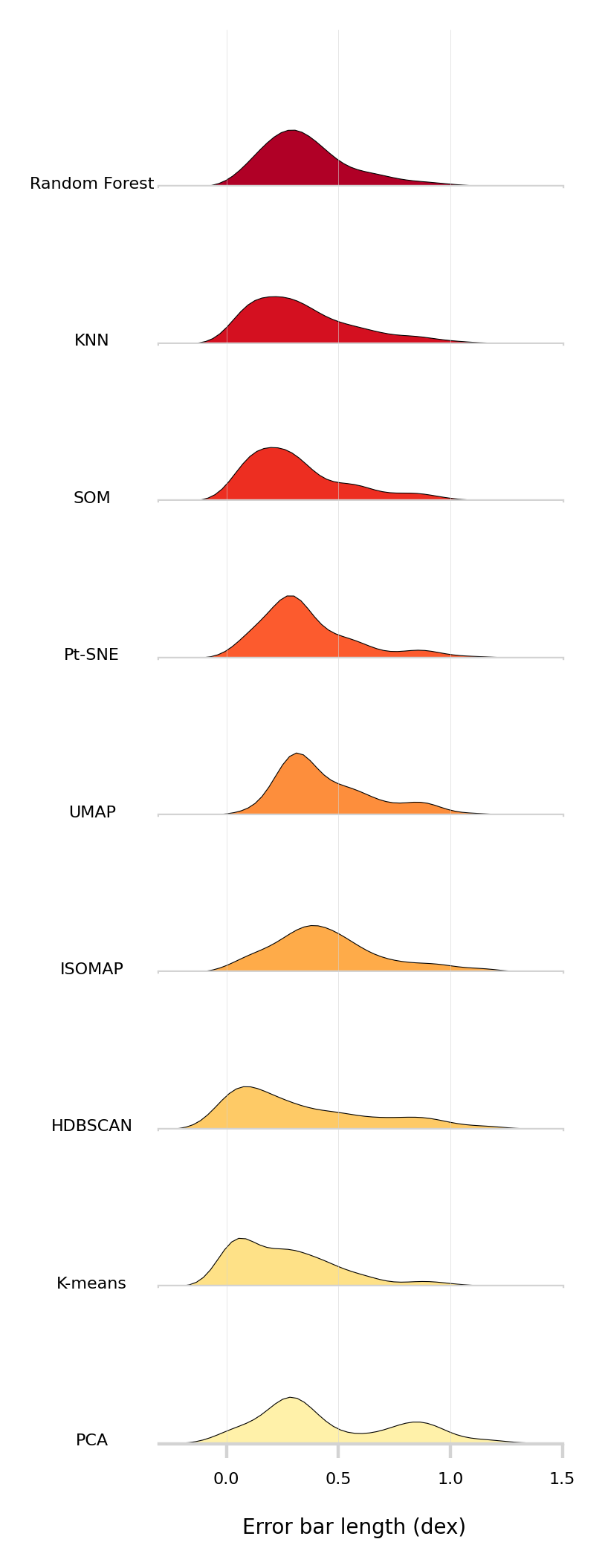}
    \caption{shows the distribution of error bar lengths for galaxy masses estimated by different ML algorithms. We analyzed 1,000 randomly selected galaxies.
    \label{fig:fig6}}
\end{figure}

\subsection{Stellar mass error bars}
We can also estimate the uncertainties associated with the stellar mass of our galaxies using different machine-learning algorithms. This approach follows the method proposed by \cite{Hemmati2019}. Here's how it works: For each galaxy, we generate 1000 realizations of its SED by randomly sampling from a distribution centered on the original photometric values, with a width defined by the photometric errors. We then use our trained ML models to predict the stellar mass for each of these 1000 SED realizations, effectively constructing the distribution of predicted masses for that galaxy across all ML algorithms. To provide a conservative and assumption-free estimate of uncertainty, especially given the presence of skewed or multi-modal distributions in our results, the minimum and maximum values of this distribution can be used to define the error bars for the stellar mass of each galaxy.

To compare the performance of our ML algorithms, we randomly selected a subset of 1,000 galaxies from our data galaxies and calculated the length of error bars (defined as the difference between the maximum and minimum values) using all considered algorithms. Figure~\ref{fig:fig6}  shows the distribution of error bar lengths for these 1,000 galaxies across all considered algorithms.

The figure reveals that PCA exhibits a wide, bimodal distribution, whereas clustering, regression, and manifold algorithms all show approximately unimodal, right-skewed distributions.

\begin{figure}
\centering
\includegraphics[width=\linewidth]{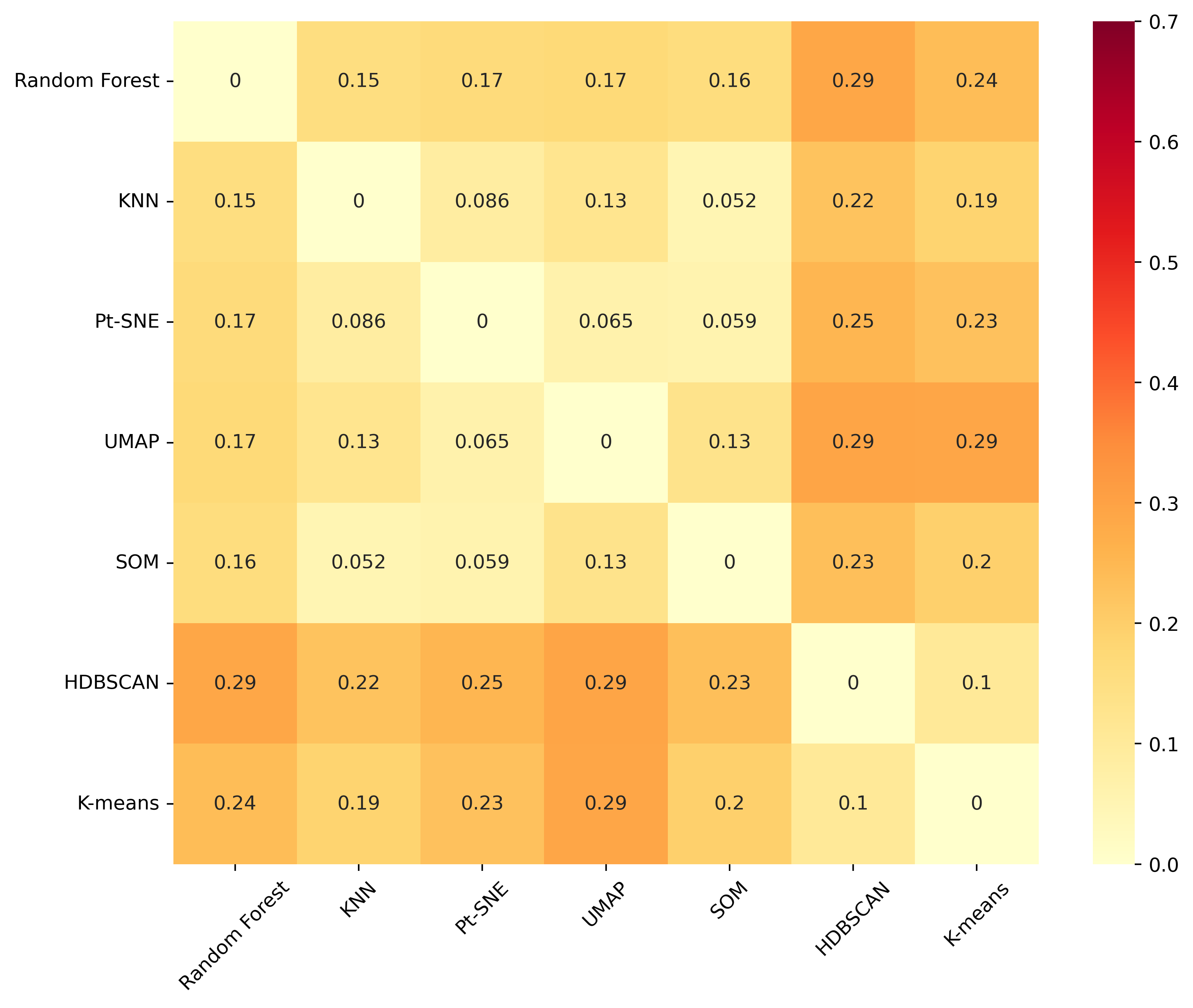}
\caption{Normalized Median Absolute Deviation (NMAD; Equation~\ref{eq:nmad}) Matrix for top-performing algorithms listed in Table~\ref{tab:tab3}.
\label{fig:fig7}}
\end{figure}

For a more detailed analysis of the similarities between the error bar length distributions produced by our algorithms, we focused on the most algorithms (excluding PCA and ISOMAP) listed in Table~\ref{tab:tab3}. For each pair of these algorithms (i, j), we computed the normalized median absolute deviation ($\sigma_{NMAD}$) using Equation~\ref{eq:nmad} and constructed a matrix, visualized in Figure~\ref{fig:fig7}. As can be seen, the maximum value of $\sigma_{NMAD}$ is less than 0.3 dex, indicating a relatively high consistency between these algorithms. This consistency is even more evident when comparing algorithms within their categories. For instance, $\sigma_{NMAD}$ for SOM and Pt-SNE is only 0.06 dex, 0.1 dex for HDBSCAN and K-means, and 0.15 dex for KNN and Random Forest.

It is worth noting that while generating 1000 realizations to estimate error bars increases the computation time for ML algorithms, the process remains significantly faster than traditional methods. For instance, repeating this process with LePhare (assuming a fitting time of 1 second per galaxy) would take $\sim 1$ million seconds ($\sim 12$ days), whereas our slowest ML algorithm (ISOMAP) completes the same task in approximately 4 hours, demonstrating a 70-fold increase in speed.

Error bar calibration analysis reveals important insights. Ideally, true galaxy masses should always fall within the estimated error bars. Using LePhare-derived masses as "true" values, we found that only $\sim 65\%$ of the masses fell within the predicted error bars for our most accurate ML algorithms. Increasing realizations to 5000 improved this near $\sim 70\%$, still significantly below expectations. This discrepancy may arise because LePhare masses are not true masses or because observational errors are underestimated.

\begin{figure}
\centering
\includegraphics[width=\linewidth]{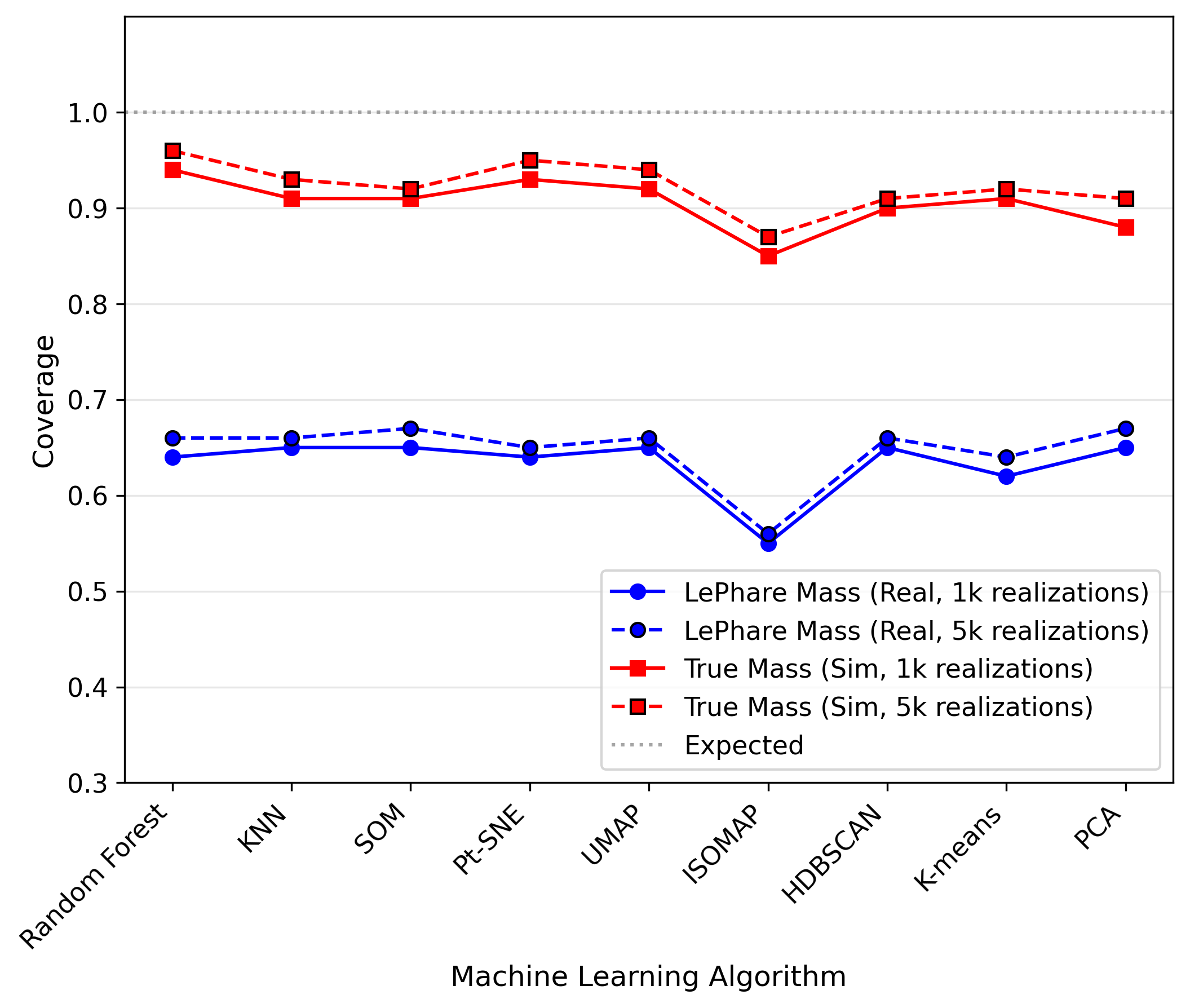}
\caption{Error bar coverage for real (blue, LePhare masses) and simulated (red, true masses) galaxy masses across ML algorithms. Simulations achieve $\gtrapprox 90\%$ coverage (proper calibration), while real data show underperformance ($\gtrapprox 65\%$). Dotted line: ideal coverage.}
\label{fig:fig8}
\end{figure}

To validate our error estimation framework, we analyzed simulated galaxies from the COSMOS2015 mock catalog \citep{laigle2019horizon}—a dataset with known stellar masses and synthetic photometry matching our observational bands (Table~\ref{tab:tab1}). Applying our BC03-trained ML models to these simulations revealed excellent error calibration: $>90\%$ of the true masses fell within the predicted error bars across approximately all ML algorithms for 1000 realizations. Increasing realizations to 5000 (e.g., for Pt-SNE) improved this to $\sim 95\%$, demonstrating proper error calibration when true masses are known. These results are summarized in Figure~\ref{fig:fig8}.

\section{Conclusion}\label{sec:6}
This study demonstrates that ML models trained on BC03 synthetic templates offer a powerful and computationally efficient alternative to traditional SED-fitting for estimating galaxy stellar masses. Through a systematic comparison of diverse ML approaches-including manifold learning, projection, clustering, and regression techniques-we show that these methods achieve mass estimates with accuracy comparable to the established LePhare SED-fitting code ($\sigma_{\rm F} \approx 0.22$ dex; see Equation~\ref{eq:rms}), while operating $10^{3}$-$10^{5}$ times faster.

Among the algorithms considered, clustering algorithms offered the best balance of accuracy and execution time. K-means performed the best overall with a $\sigma_{F}$ of 0.19 dex and a speedup of approximately $1.5 \times 10^{5}$ times compared to LePhare. HDBSCAN also demonstrated better accuracy with a $\sigma_{F}$ of 0.21 dex and a speedup of $6.9 \times 10^{4}$.

In the manifold learning category, all algorithms except ISOMAP showed promising results in accuracy and fitting time. SOM stood out in mapping speed, particularly when dealing with a large number of galaxies, achieving speedups of $1.2 \times 10^{4}$. ISOMAP, however, had a higher $\sigma_{F}$ of 0.27 dex and a lower speedup of $5.5 \times 10^{1}$.

For regression algorithms, both Random Forest and KNN demonstrated strong performance, achieving $\sigma_{F}$s values of 0.22 dex and 0.21 dex respectively while maintaining computational efficiency. These methods delivered speed improvements of $1.9 \times 10^{4}$ and $2.1 \times 10^{4}$ compared to traditional SED-fitting with LePhare.

While PCA delivered the fastest processing ($2.2 \times 10^{5}$ speedup), its accuracy was the lowest among all methods ($\sigma_{F}$ = 0.30 dex)

Furthermore, we demonstrated the use of ML algorithms to estimate error bars associated with stellar mass estimates. Overall, our analysis highlights a relatively good level of consistency between the most accurate algorithms ($\sigma_{NMAD}<0.30$ dex), particularly within their respective categories ($\sigma_{NMAD}<0.15$ dex).

Our selection of ML algorithms prioritized prominent choices from various categories. We favored algorithms with stable performance using default settings whenever possible. This approach ensures reproducibility while acknowledging that other algorithms could be explored in future studies.

In this study, we used stellar masses derived from LePhare as a reference point to evaluate the performance of various ML algorithms. While LePhare is a commonly used method, it's important to acknowledge that the "true" stellar mass values for these galaxies are unknown. For SFR, \cite{Davidzon2022} reported that SOM-derived SFRs were more consistent with independent measurements obtained through UV-to-FIR photometry \citep{Barro2019} compared to LePhare SFRs. This suggests that ML approaches may offer advantages in accuracy for the physical properties of galaxies. Future investigations using cosmological simulations or well-calibrated SPS models could provide a valuable tool for directly comparing the accuracy of LePhare and different ML algorithms for stellar mass estimation.

In conclusion, this study demonstrates the effectiveness of ML algorithms for estimating stellar masses, achieving accuracy comparable to traditional methods like LePhare at significantly faster speeds. This substantial speedup paves the way for analyzing upcoming large-scale astronomical surveys.

\bibliography{ref}

\end{document}

%% file: band_table.tex
\begin{table}[h]
\caption{Photometric bands from the COSMOS2015 survey \citep{Laigle2016} used in this study.}
\setlength{\belowcaptionskip}{10pt}
\centering
\begin{tabular}{ccccc}
\hline \hline 
\toprule
Instrument                & Filter  & Central                 & Width     & $3\sigma$ depth \\
/Telescope                &         &$\lambda (\AA)$          & $(\AA)$   & ($3^{"}/2^{"}$) \\
(Survey)                  &         &                         &           & $\pm$ 1         \\
\midrule
MegaCam/CFHT              & u*      & 3823.3                  & 670       & 26.6/27.2        \\
\midrule
Suprime-Cam               & B       & 4458.3                  & 946       & 27.0/27.6        \\
/Subaru                   & V       & 5477.8                  & 955       & 26.2/26.9        \\
                          & $r$     & 6288.7                  & 1382      & 26.5/27.0        \\
                          & $i^{+}$ & 7683.9                  & 1497      & 26.2/26.9         \\
                          & $z^{++}$& 9105.7                  & 1370      & 25.9/26.4         \\
\midrule
VIRCAM                    & Y       & 10214.2                 & 970       & 25.3/25.8         \\
/UltraVISTA               & J       & 12534.6                 & 1720      & 24.9/25.4         \\
                          & H       & 16453.4                 & 2900      & 24.6/25.0         \\
                          & $K_{s}$ & 21539.9                 & 3090      & 24.7/25.2         \\
\midrule
IRAC/\textit{Spitzer}     & ch1     & 35634.3                 & 7460      & 25.5               \\
                          & ch2     & 45110.1                 & 10110     & 25.5               \\
\bottomrule
\label{tab:tab1}
\end{tabular}
\end{table}

%% file: uni_table0.tex
\begin{table}
\centering
\caption{Unsupervised and Supervised learning Algorithms Considered in this study.}
\label{tab:combined_algorithms}
\setlength{\belowcaptionskip}{10pt}
\begin{tabular}{c|ccc}
\hline \hline 
\toprule
\multicolumn{1}{c|}{Kind} & Category & Algorithm & Open-source \\
\midrule
\multirow{6}{*}{Unsupervised} & Manifold & Pt-SNE  & \href{https://opentsne.readthedocs.io/en/latest/parameters.html}{openTSNE} \\
\multirow{6}{*}{Learning} & Learning & UMAP     & \href{https://umap-learn.readthedocs.io/en/latest/}{umap} \\
&  & SOM      & \href{https://github.com/ldocao/sompy/}{sompy} \\
&  & ISOMAP   & \href{https://scikit-learn.org/stable/modules/generated/sklearn.manifold.Isomap.html}{scikit-learn} \\ \cline{2-4}
& Projection & PCA      & \href{https://scikit-learn.org/stable/modules/generated/sklearn.decomposition.PCA.html}{scikit-learn} \\\cline{2-4}
& Clustering & K-means   & \href{https://scikit-learn.org/stable/modules/generated/sklearn.cluster.KMeans.html}{scikit-learn} \\
&  & HDBSCAN   & \href{https://scikit-learn.org/stable/modules/generated/sklearn.cluster.HDBSCAN.html}{scikit-learn} \\ \midrule
\multirow{1}{*}{Supervised} & Regression  & Random Forest & \href{https://scikit-learn.org/stable/modules/generated/sklearn.ensemble.RandomForestRegressor.html}{scikit-learn} \\
\multirow{1}{*}{Learning} &   & KNN          & \href{https://scikit-learn.org/stable/modules/generated/sklearn.neighbors.KNeighborsRegressor.html}{scikit-learn} \\
\bottomrule
\end{tabular}
\label{tab:tab2}
\end{table}

%% file: uni_table.tex
\begin{table*}
\centering
\begin{tabular}{c|ccccccc}
\hline \hline 
\toprule
\multicolumn{1}{c|}{Kind} & Category & Algorithm & OLF & $\sigma_{F}$ & $\sigma_{NMAD}$ & $\sigma_{0}$ & $t_{LePhare}/t_{Algorithm}$ \\ 
\midrule
\multirow{7}{*}{Unsupervised Learning} & Clustering & K-means & 0.010 & 0.190 & 0.180 & 0.180 & $\sim 1.5 \times 10^{5}$ \\
&  & HDBSCAN & 0.021 & 0.213 & 0.202 & 0.192 & $\sim 6.9 \times 10^{4}$ \\ \cline{2-8}
& Manifold & Pt-SNE & 0.018 & 0.193 & 0.186 & 0.173 & $\sim 3.2 \times 10^{3}$ \\
& Learning & SOM ($60 \times 80$) & 0.020 & 0.202 & 0.182 & 0.181 & $\sim 1.2 \times 10^{4}$ \\
&  & UMAP & 0.020 & 0.201 & 0.191 & 0.180 & $\sim 2.1 \times 10^{3}$ \\
&  & ISOMAP & 0.053 & 0.268 & 0.281 & 0.230 & $\sim 5.5 \times 10^{1}$ \\ \cline{2-8}
& Projection & PCA & 0.111 & 0.303 & 0.277 & 0.225 & $\sim 2.2 \times 10^{5}$ \\ \midrule
\multirow{2}{*}{Supervised Learning} & Regression & KNN & 0.020 & 0.206 & 0.187 & 0.185 & $\sim 2.1 \times 10^{4}$ \\
&  & Random Forest & 0.022 & 0.220 & 0.215 & 0.200 & $\sim 1.9 \times 10^{4}$ \\

\bottomrule
\end{tabular}

\caption{compares galaxy masses derived from ML algorithms with those derived from LePhare using outlier fraction and variations for approximately 14,000 galaxies. Additionally, it summarizes the average speedup achieved by the machine learning algorithms compared to LePhare.}
\label{tab:tab3}
\end{table*}